\documentclass[cernpreprint, atlasdraft=false, coverpage=false texlive=2023, UKenglish, texmf, orcidlogo, texmf, orcidlogo]{atlasdoc}
\usepackage{atlaspackage}
\usepackage{atlasbiblatex}

\usepackage{atlasphysics}
\graphicspath{{logos/}{figures/}}

\usepackage{hyperref}

\newcommand{\sqn}{\sqrt{s_{_{\mathrm{NN}}}}}
\newcommand{\Nch}{\textit{N}_\mathrm{ch}}
\newcommand{\Nchn}{(\textit{N}_\mathrm{ch})^{n-1}}

\newcommand{\NchR}{\textit{N}_\mathrm{ch}^\mathrm{rec}}

\newcommand{\SumET}{\Sigma E_\mathrm{T}}

\newcommand{\lr}[1]{\left\langle #1\right\rangle}

\newcommand*{\HIJING}{\textsc{Hijing}\xspace}
\AtlasTitle{Disentangling sources of momentum fluctuations in Xe+Xe and Pb+Pb collisions with the ATLAS detector}

\AtlasAbstract{
High-energy nuclear collisions create a quark-gluon plasma, whose initial condition and subsequent expansion vary from event to event, impacting the distribution of the event-wise average transverse momentum ($P([p_{\mathrm{T}}])$). Distinguishing between contributions from fluctuations in the size of the nuclear overlap area (geometrical component) and other sources at fixed size (intrinsic component) presents a challenge. Here, these two components are distinguished by measuring the mean, variance, and skewness of $P([p_{\mathrm{T}}])$ in $^{208}$Pb+$^{208}$Pb and $^{129}$Xe+$^{129}$Xe collisions at $\sqrt{s_{{\mathrm{NN}}}} = 5.02$ and 5.44 TeV, respectively, using the ATLAS detector at the LHC. All observables show distinct changes in behavior in ultra-central collisions, where the geometrical variations are suppressed as the overlap area reaches its maximum. These results demonstrate a new technique to disentangle geometrical and intrinsic fluctuations, enabling constraints on initial condition and properties of the quark-gluon plasma, such as the speed of sound.
}

\AtlasRefCode{HION-2021-20}

\PreprintIdNumber{CERN-EP-2024-173}

\AtlasJournalRef{\PRL 133 252301 (2024)}
\AtlasDOI{https://doi.org/10.1103/PhysRevLett.133.252301}

\AtlasTitle{Disentangling sources of momentum fluctuations in Xe+Xe and Pb+Pb collisions with the ATLAS detector}
\AtlasAbstract{High-energy nuclear collisions create a quark-gluon plasma, whose initial condition and subsequent expansion vary from event to event, impacting the distribution of the event-wise average transverse momentum ($P([p_{\mathrm{T}}])$). Disentangling the contributions from fluctuations in the nuclear overlap size (geometrical component) and other sources at a fixed size (intrinsic component) remains a challenge. This problem is addressed by measuring the mean, variance, and skewness of $P([p_{\mathrm{T}}])$ in $^{208}$Pb+$^{208}$Pb and $^{129}$Xe+$^{129}$Xe collisions at $\sqrt{s_{{\mathrm{NN}}}} = 5.02$ and 5.44 TeV, respectively, using the ATLAS detector at the LHC. All observables show distinct features in ultra-central collisions, which are explained by a suppression of the geometrical component as the overlap area reaches its maximum. These results demonstrate a new technique to separate geometrical and intrinsic fluctuations, providing constraints on initial conditions and properties of the quark-gluon plasma, such as the speed of sound.}
\AtlasRefCode{HION-2021-20}
\PreprintIdNumber{CERN-EP-2024-173}
\AtlasJournal{Physical Review Letters}
\begin{document}
\maketitle

High energy nuclear collisions at the Relativistic Heavy Ion Collider (RHIC) and the Large Hadron Collider (LHC) create a strongly-interacting state of matter known as quark-gluon plasma (QGP)~\cite{Busza:2018rrf}. The hydrodynamic expansion of the QGP induces a significant boost to the transverse momentum ($\pT$) of the final-state particles. This boost transforms the initial shape anisotropies and size variations of the QGP into final state anisotropic flow~\cite{Teaney:2010vd, Gardim:2011xv, Niemi:2012aj} and variations in the event-wise average transverse momentum, $[\pT]$~\cite{Bozek:2012fw}. Comparisons of anisotropic flow with model calculations have provided crucial insights into the initial conditions, such as the overlap area and nucleon/subnucleonic fluctuations, as well as transport properties of the QGP, such as shear and bulk viscosities~\cite{Romatschke:2017ejr, Busza:2018rrf}. However, quantitative extraction of these properties has significant uncertainties due to limited knowledge about the initial conditions~\cite{Bernhard:2019bmu, Nijs:2020roc}.

Naturally, progress can be achieved by studying the distribution $P([\pT])$ for events with similar impact parameters. Since the $[\pT]$ is sensitive to the radial expansion or radial flow in each event, $P([\pT])$ provides insights into the initial state, the equation of state (EOS), and the associated speed-of-sound squared ($c_s^2$) in the QGP~\cite{Gardim:2019xjs,Gardim:2019brr,Samanta:2023amp,Samanta:2023kfk,Nijs:2023bzv, Gardim:2024zvi,SoaresRocha:2024drz}. The distribution $P([\pT])$ can be characterized through its moments: the mean $\lr{[\pT]}$, variance $\lr{(\delta \pT)^2}$, and skewness $\lr{(\delta \pT)^3}$, where $\delta \pT = [\pT]-\lr{[\pT]}$, and "$\lr{}$" denotes an ensemble average.

Most sources contributing to $P([\pT])$ appear stochastic, including fluctuations in the transverse size $R$ of the overlap region, the positions of nucleons and partons in the initial state, energy deposition, and the temperature of the QGP fluid. These sources can be categorized into ``geometrical fluctuations'' that reflect the hydrodynamic response to event-by-event variations in $R$ with $\delta\pT/\lr{[\pT]} \approx -\delta R/\lr{R}$~\cite{Bozek:2012fw}, and ``intrinsic fluctuations'' that include other sources of $\delta\pT$ at a fixed $R$~\cite{Samanta:2023amp}. If nuclear collisions are considered as a superposition of independent particle production from participating nucleons, followed by final state interactions, both geometrical and intrinsic fluctuations are expected to approximately scale with the charged particle multiplicity ($\Nch$): $\lr{(\delta \pT)^2}\propto 1/\Nch$ and $\lr{(\delta \pT)^3} \propto 1/\Nch^{\;2}$. This is known as the independent superposition scenario~\cite{Gavin:2011gr, Cody:2021cja}.

These two components of fluctuations can be investigated using moments of $P([\pT])$ in ultra-central collisions (UCC)~\cite{Gazdzicki:1992ri,Samanta:2023amp, Samanta:2023kfk}. As the impact parameter approaches zero in UCC, $R$ reaches its maximum value, suppressing geometrical fluctuations and leading to a deviation from the anticipated $1/\Nch$ scaling. In contrast, intrinsic fluctuations continue to follow the expected scaling behavior. This interplay between the geometrical and intrinsic fluctuations gives rise to complex behaviors in the moments of $P([\pT])$. Measurements in UCC can constrain the properties of the intrinsic fluctuations, which are sensitive to $c_s^2$~\cite{Gardim:2019xjs}.

The dependence of the mean and variance of $P([\pT])$ on multiplicity has been measured across various system sizes and collision energies~\cite{NA49:1999inh, PHENIX:2002aqz, PHENIX:2003ccl, STAR:2003cbv, NA49:2003hxt, CERES:2003sap, STAR:2005vxr, NA49:2008fag,STAR:2013sov, ALICE:2014gvd, STAR:2019dow}. These studies show an increase in $\lr{[\pT]}$ towards more central collisions, with the variance following the expected power-law scaling. Recently, ALICE reported measurements of skewness and kurtosis in Xe+Xe and Pb+Pb collisions~\cite{ALICE:2023tej}, though only within broad multiplicity ranges. CMS performed a detailed study of the behavior of $\lr{[\pT]}$ in Pb+Pb UCC, claiming to extract the $c_s^2$~\cite{CMS:2024sgx} but with caveats~\cite{Nijs:2023bzv, Gardim:2024zvi,SoaresRocha:2024drz}. These measurements could not disentangle the geometrical and intrinsic components of $P([\pT])$. To achieve this goal, a precise measurement of higher-order moments in UCC is necessary.

This Letter reports the measurement of the mean, variance, and skewness of $P([\pT])$ as a function of $\Nch$ in $^{208}$Pb+$^{208}$Pb collisions at $\sqn = 5.02$~TeV and $^{129}$Xe+$^{129}$Xe collisions at $\sqn = 5.44$~TeV. A smaller transverse size and multiplicity range in Xe+Xe than Pb+Pb provide a unique opportunity to study the role of system size in the scaling behavior of these moments.

The measurements are performed using the ATLAS inner detector~(ID), forward calorimeter~(FCal), and zero-degree calorimeters~(ZDCs) along with the trigger and data acquisition systems~\cite{Aad:2008zzm, Aad:ID,Aaboud:2016leb}. The ID detects charged particles within $|\eta| < 2.5$\footnote{ATLAS uses a right-handed coordinate system with its origin at the nominal IP in the center of the detector and the $z$-axis along the beam pipe. The $x$-axis points from the IP to the center of the LHC ring, and the $y$-axis points upward. Cylindrical coordinates $(r,\phi)$ are used in the transverse plane, $\phi$ being the azimuthal angle around the beam pipe. The pseudorapidity is defined in terms of the polar angle $\theta$ as $\eta=-\ln\tan(\theta/2)$.} using a combination of silicon pixel and microstrip detectors, along with a straw-tube transition-radiation tracker, all immersed in a 2T axial magnetic field~\cite{Aad:ID}. The FCal consists of three sampling layers, covering $3.2<|\eta|< 4.9$. The ZDCs are positioned at $\pm$140~m from the interaction point (IP), and detect neutrons with $|\eta|>8.3$. The ATLAS trigger system~\cite{Aaboud:2016leb} consists of a hardware-based level-1 (L1) trigger and a software-based high-level trigger. A software suite~\cite{SOFT-2022-02} is used in data simulation, in the reconstruction and analysis of real and simulated data, in the detector operations, and in the trigger and data acquisition systems.

This analysis uses 470~$\mu\mathrm{b}^{-1}$ of Pb+Pb data collected in 2015 and 3~$\mu\mathrm{b}^{-1}$ of Xe+Xe data collected in 2017. Pb+Pb events are selected by requiring the total transverse energy deposited in calorimeters over $|\eta| < 4.9$ at L1 ($\eT^{L1}$) to exceed 50 GeV. Additionally, dedicated central collision triggers are used to enhance the number of events with large FCal transverse energy~\cite{ATLAS:2019peb}. Xe+Xe events are selected by requiring $\eT^{L1}>4$ GeV.

Charged-particle tracks are reconstructed from hits in the ID using a reconstruction and selection procedure optimized for heavy-ion collisions~\cite{Aad:2015wga}. Tracks used in this analysis must have $\pT>0.5$~GeV and $|\eta|<2.5$, with the total number of such tracks in each event denoted by $\NchR$. Events containing multiple inelastic collisions (pileup) are suppressed by exploiting the correlation between $\NchR$ and the transverse energy measured in the FCal, $\SumET$. The pileup probability is 0.17\% in Pb+Pb collisions and a factor of ten smaller in Xe+Xe collisions. In the Pb+Pb dataset, additional pileup suppression is achieved by exploiting the correlation between the energy deposited in the ZDCs and $\SumET$~\cite{ATLAS:2022kqu}. The residual pileup fraction is less than 0.01\% in central collisions (see Appendix).

Events are categorized into centrality intervals using a Glauber model~\cite{Miller:2007ri} parameterization of the $\SumET$ distribution~\cite{ATLAS:2019peb}. Each interval represents a range in $\SumET$, starting at 0\% for the most central collisions with the highest $\SumET$ value and ending at 80\%. In this analysis, events within the top 5\% centrality, where observables deviate strongly from power-law scaling, are denoted as UCC. These events correspond to $\SumET > $ 3.62~TeV for Pb+Pb and 2.27~TeV for Xe+Xe collisions, respectively.

The track reconstruction efficiency, $\epsilon(\pT,\eta,\NchR)$, is evaluated using Monte Carlo (MC) simulated events for Pb+Pb and Xe+Xe collisions generated with $\HIJING$~\cite{Gyulassy:1994ew}. For this evaluation, charged particles are defined according to Ref.~\cite{Aad:2015wga}. The detector response is simulated using $\GEANT$~\cite{Agostinelli:2002hh, Aad:2010ah}, and events are reconstructed with the same algorithms used for the data. For charged particles with $p_{\mathrm{T}}>0.8$~GeV, where the efficiency varies slowly, the efficiency in UCC Pb+Pb collisions ranges from $71 \%$ at $\eta \approx 0$ to about $40 \%$ for $|\eta|>2$. The efficiency decreases by $12 \%$ from 0.8 GeV to 0.5~GeV, averaged over the full $\eta$ range. In peripheral collisions, the efficiency is up to 4\% higher. The rate of falsely reconstructed (``fake'') tracks, $f(\pT,\eta,\NchR)$, is significant for $p_{\mathrm{T}}<1$~GeV in UCC collisions, ranging from $2 \%$ for $|\eta|<1$ to $8 \%$ at larger $|\eta|$. The fake-track rate drops rapidly for higher $p_{\mathrm{T}}$ and more peripheral collisions. Within the $\NchR$ range covered by UCC events, efficiency drops by 1\% with increasing $\NchR$, while the fake rate increases by 4\%. At the same $\NchR$, the efficiency in Xe+Xe is about 2\% lower than in Pb+Pb, while fake rates agree within 1\%.

The moments of $P([\pT])$ are calculated using computational methods similar to those developed for the anisotropic flow~\cite{Bilandzic:2010jr, Bhatta:2021qfk}. The $[\pT]$ and $n$-particle correlators in a single event are computed as \mbox{$[\pT]\!=\!\sum_{i}\!w_{i}p_{i}/\sum_{i}\! w_{i}$}, $c_{2}\!=\!\sum_{i\neq j}\!w_{i}w_{j}\delta p_{i}\delta p_{j}/\sum_{i\neq j }\!w_{i}w_{j}$ and $c_{3}\!=\!\sum_{i\neq j \neq k }\!w_{i}w_{j}w_{k}\delta p_{i}\delta p_{j}\delta p_{k}\!/\!\sum_{i\neq j\neq k }\!w_{i}w_{j}w_{k}$. Here, \mbox{$\delta p_{i} \equiv p_{\mathrm{T},i}- \lr{[\pT]}$}, and $w_{i}$ are weights applied to track $i$ to correct for reconstruction efficiency $\epsilon_i$ and fake rate $f_i$: $w_i \equiv (1-f_i)/\epsilon_i$~\cite{Aad:2019fgl}. This track-by-track weighting procedure is sufficient since the migrations in $\eta$, $\phi$, and $\pT$ between the truth and reconstructed tracks are minimal~\cite{ATLAS:2022kqu}. The $n^{\mathrm{th}}$ central moment of the corresponding $P([\pT])$ is obtained by averaging $c_{n}$ over a given event ensemble in unit $\NchR$ intervals, denoted as $\lr{c_{n}}= \lr{(\delta \pT )^n}$. The $\Nch$ is calculated as $\Nch = \sum_{i} w_{i}$, using tracks in $0.5 <\pT< 5$ GeV and $|\eta|<2.5$.

The variance and skewness are normalized into dimensionless quantities:
\begin{align}
& k_{2}=\frac{\lr{c_{2}}}{\lr{[\pT]}^{2}},\; k_{3}=\frac{\lr{c_{3}}}{\lr{[\pT]}^{3}},\; \gamma = \frac{\lr{c_{3}}}{\lr{c_{2}}^{3/2}} ,\; \Gamma = \frac{\lr{c_{3}}\lr{[\pT]}}{\lr{c_{2}}^{2}}. \label{eq:ck2}
\end{align}
The ``standard skewness'', $\gamma$, corresponds to the skewness for a distribution with unit variance, while $\Gamma$ is referred to as the ``intensive skewness''~\cite{Giacalone:2020lbm}. Statistical uncertainties for these observables are computed using a Poisson bootstrap method~\cite{ATLAS:2021kho}. In the independent superposition scenario, $k_{2} \propto 1/\Nch$, $k_{3}\propto 1/(\Nch)^2$, $\gamma \propto 1/\sqrt{\Nch}$, while $\Gamma$ should be roughly independent of $\Nch$.

Systematic uncertainties stem from track selection, reconstruction efficiency, residual pileup, centrality definition, and MC consistency check. Their values in the 0--60\% centrality range are summarized as follows. Track selection uncertainties are assessed by comparing nominal results against those obtained under stricter criteria, resulting in deviations of $<0.5$\% for $\lr{[\pT]}$, 0.5--3\% for $k_{2}$, 0--1.5\% for $k_{3}$, 0.5--4\% for $\gamma$, and 0.5--1.5\% for $\Gamma$. Uncertainties in the efficiency, partially due to modeling of the detector material in $\GEANT$, can be as large as 4\%~\cite{Aad:2019fgl}. The impact of the efficiency uncertainties on the measured moments are around 1\% for $\lr{[\pT]}$, 0.5\% for $k_{2}$, 2--2.5\% for $k_{3}$, 1--1.5\% for $\gamma$, and 1.5--2.5\% for $\Gamma$. The impact of the azimuthal efficiency variation is assessed by including it in the track weight; no observable changes are found in the results. Residual pileup effects are estimated by adjusting the pileup rejection criteria, leading to uncertainties less than 0.5\% for all observables. Centrality definition uncertainties, accessed by varying the Glauber model parameters, are relevant only when results are presented in centrality intervals and are below 0.5\% in UCC for all observables.

The $\HIJING$ MC samples are used to evaluate the consistency of the $[\pT]$-moments, obtained using truth particles or the reconstructed tracks with the same correction procedures for the real data applied~\cite{ATLAS:2019peb,ATLAS:2019dct}~\footnote{Although the $[\pT]$-moments in $\HIJING$ differ from those in data, this study primarily checks if the analysis procedure applied to real data can reproduce $[\pT]$-moments at the truth level.} This check also accounts for the smearing of measured $[\pT]$ and $\Nch$ from the true values. The differences are less than $0.25\%$ for $\lr{[\pT]}$ and $k_{2}$, and are about $1.2\%$ for $k_{3}$, $\gamma$ and $\Gamma$.

\begin{figure}[h!]
\centering
\includegraphics[width=0.9\linewidth]{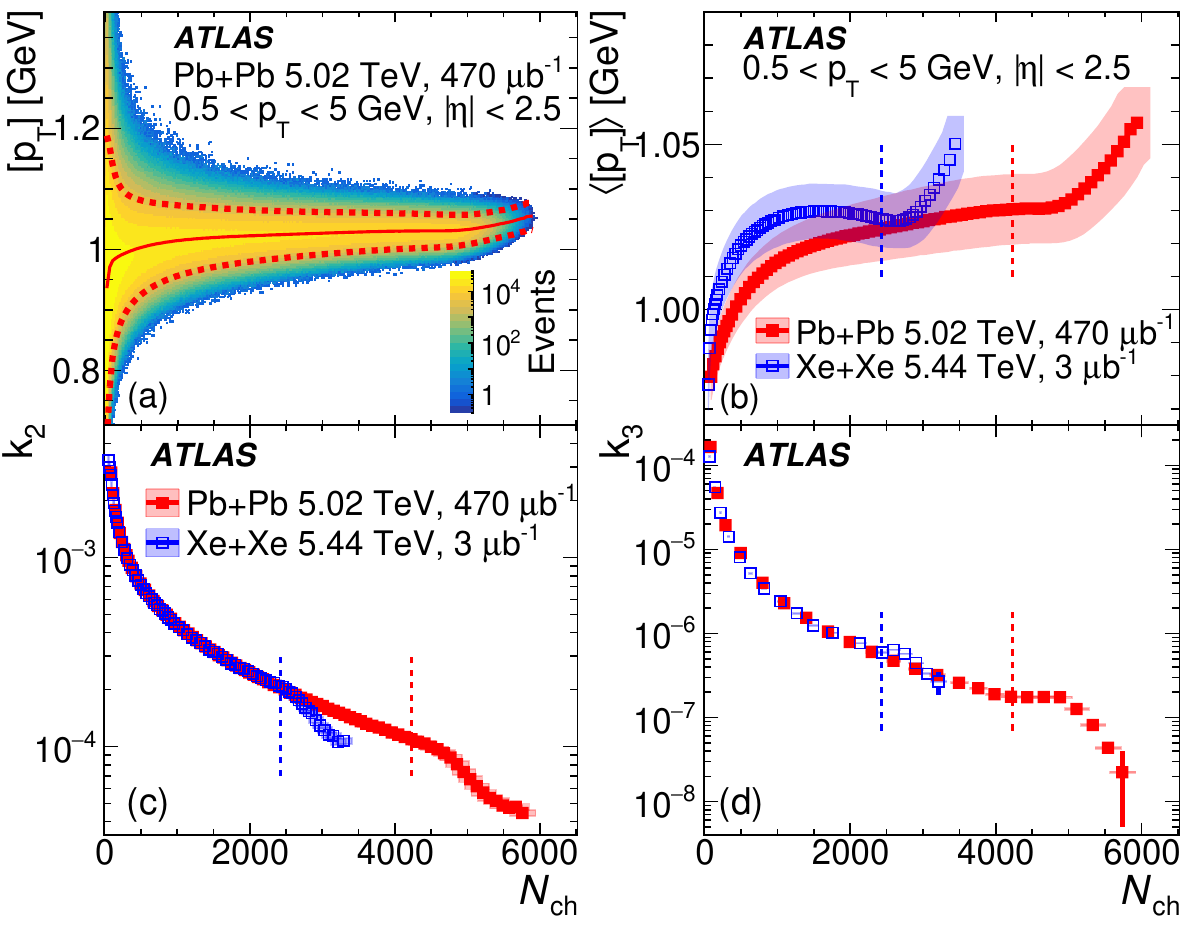}
\caption{ Panel (a) depicts the 2D distribution of $[\pT]$ versus $\Nch$ in Pb+Pb collisions, where the solid and dashed lines indicate the mean and two standard deviations, respectively. Panels (b), (c), and (d) show the $\Nch$ dependence of $\lr{[\pT]}$, $k_2$, and $k_3$, respectively. The error bars represent statistical uncertainties of the measurement, whereas the shaded boxes represent the systematic uncertainties in both $x$- and $y$-axes. The vertical dashed lines mark the $\Nch$ values 4230 and 2425, corresponding to 5\% centrality in Pb+Pb and Xe+Xe collisions, respectively.}
\label{fig:fig1}
\end{figure}

Total systematic uncertainties for each observable are obtained by adding the individual sources in quadrature. Track selection is the dominant source of systematic uncertainty, particularly in mid-central and central collisions. The total uncertainties are less than 1\% for $\lr{[\pT]}$, 2--4\% for $k_{2}$, 2--5\% for $k_{3}$, and 2--4\% for $\gamma$ and $\Gamma$ in both systems; they are smaller than the statistical uncertainties except for $\lr{[\pT]}$. The uncertainty for $\Nch$ is dominated by the correction on tracking efficiency and fake tracks, and reaches up to 3\% in Pb+Pb UCC.

The two-dimensional (2D) distribution of $[\pT]$ versus $\Nch$ is shown in Figure~\ref{fig:fig1}(a) for Pb+Pb collisions, whose mean and widths at fixed $\Nch$ are indicated by the solid and dashed lines, respectively. The data show a mild increase of the means and a narrowing of the widths with increasing $\Nch$.

Figure~\ref{fig:fig1}(b)-(d) present the measured moments in Pb+Pb and Xe+Xe collisions. An increase of $\lr{[\pT]}$ with $\Nch$ is observed in peripheral collisions, which weakens in mid-central collisions. The values of $k_2$ and $k_3$ show a power-law-like decrease with increasing $\Nch$. In UCC, all three observables deviate sharply from their mid-central trends: $\lr{[\pT]}$ increases while $k_2$ and $k_3$ decrease towards higher $\Nch$ values.

\begin{figure}[h!]
\centering
\includegraphics[width=0.5\linewidth]{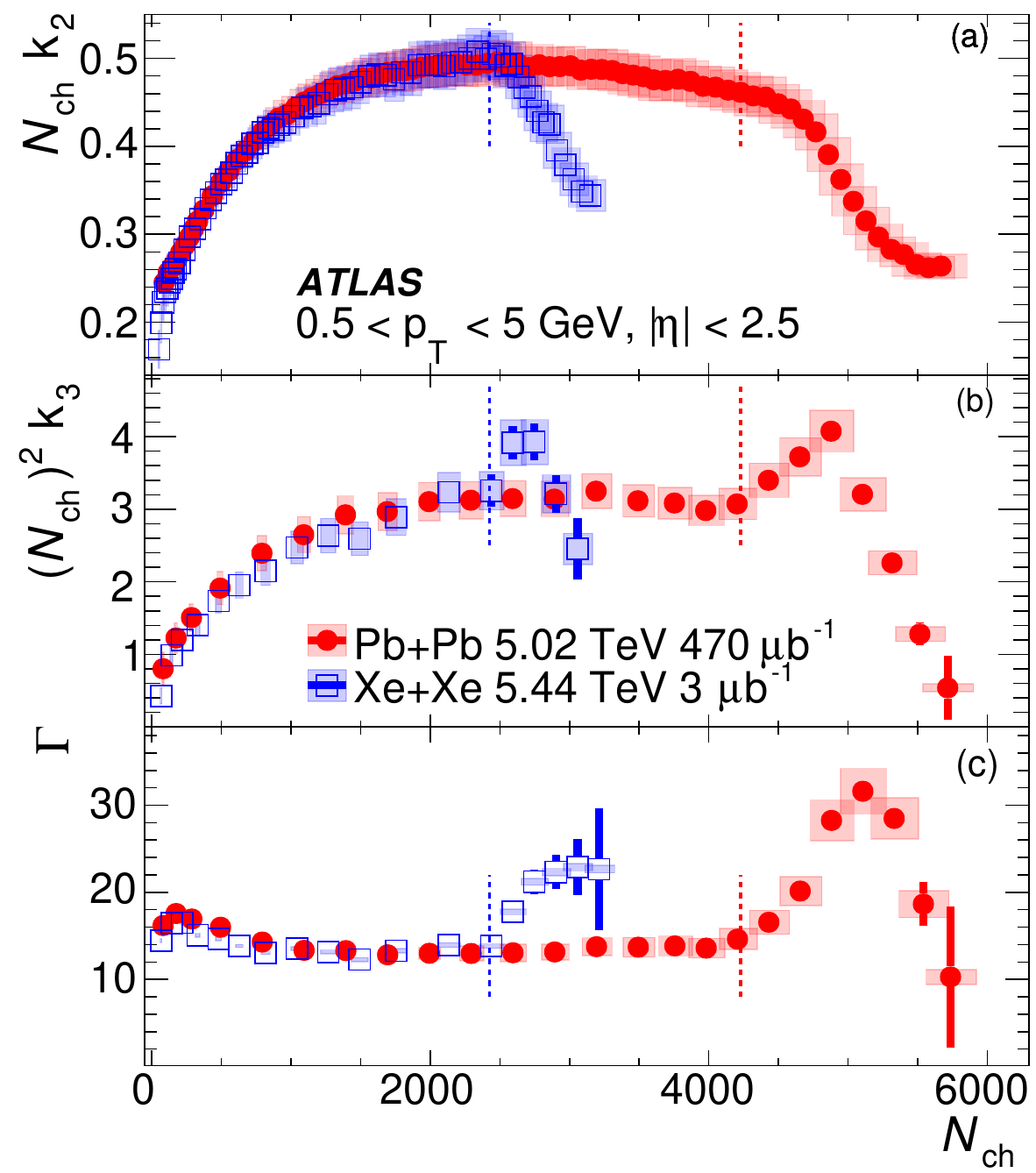}
\caption{ The values of (a) $\Nch k_{2}$, (b) $(\Nch)^{2} k_{3}$, and (c) $\Gamma$, as a function of $\Nch$ in Pb+Pb and Xe+Xe collisions. The error bars represent the statistical uncertainties of the measurement, whereas the shaded boxes represent the systematic uncertainties of the data along the $x$- and $y$-axes. The vertical dashed lines mark the $\Nch$ values corresponding to 5\% centrality in Pb+Pb and Xe+Xe collisions, respectively.}
\label{fig:fig2}
\end{figure}

Figure~\ref{fig:fig2} displays $\Nch$ dependence of $\Nchn k_n$ and $\Gamma$, which quantify any deviations from the expected power-law scaling behavior. Increases of $\Nch k_2$ and $(\Nch)^2 k_3$ with $\Nch$, consistent with the onset of radial flow~\cite{Voloshin:2003ud}, are observed up to $\Nch\approx 1500$ in Pb+Pb and Xe+Xe collisions. Beyond this range, both observables vary more gradually until the UCC region.  Meanwhile, $\Gamma$ decreases slightly from peripheral to mid-central collisions, remaining flat until the UCC region.

In UCC, $\Nch k_2$ decreases significantly while $(\Nch)^2 k_3$ and $\Gamma$ show abrupt increases followed by sharp decreases. These non-monotonic trends are consistent with the expected suppression of the distribution of $R$, $P(R)$, on the larger $R$ side~\cite{Samanta:2023amp}. This suppression initially leads to a positive skew, which eventually vanishes as the variance of $P(R)$ approaches zero.

\begin{figure}[h!]
\centering
\includegraphics[width=0.95\linewidth]{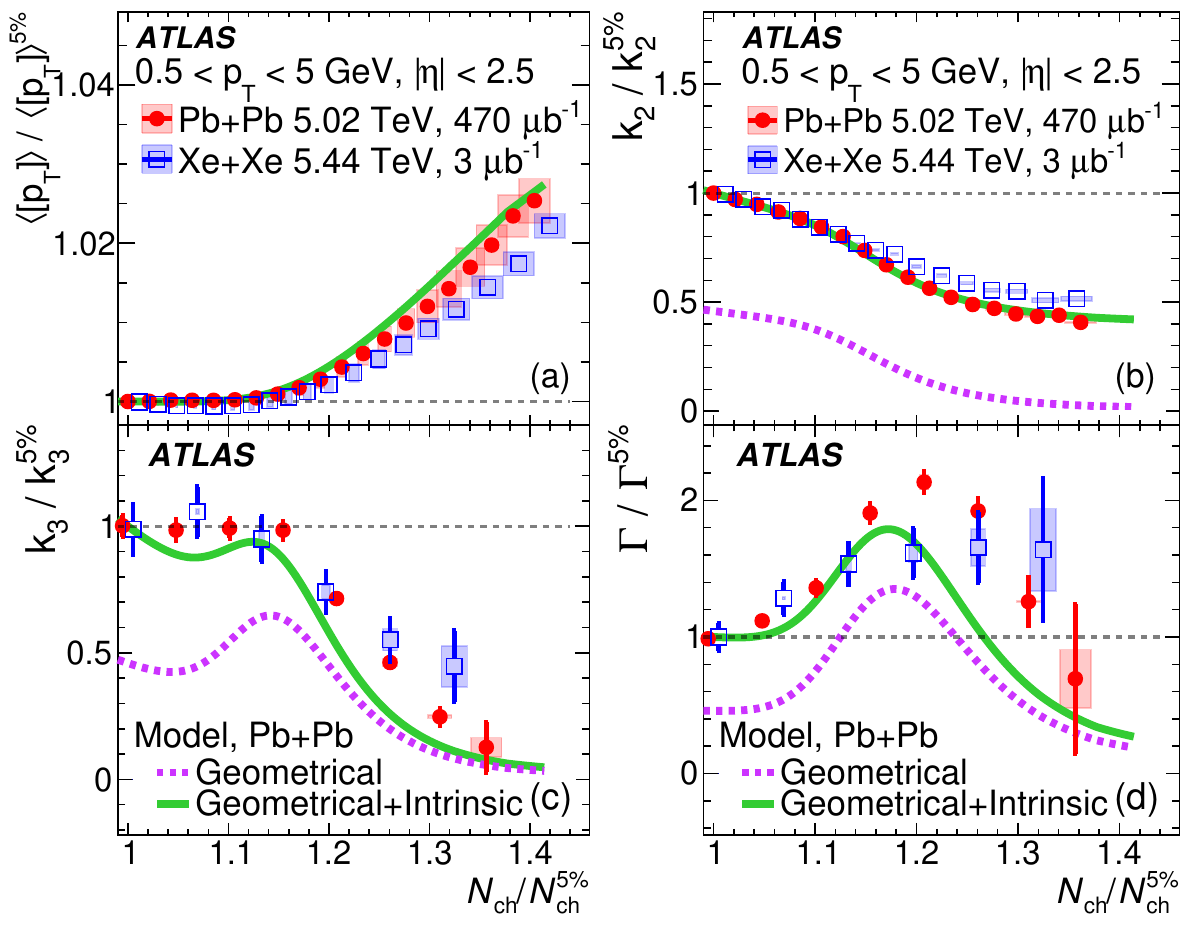}
\caption{ The (a) $\lr{[\pT]}$, (b) $k_2$, (c) $k_3$ and (d) $\Gamma$, scaled by their values at 5\% centrality as a function of $\Nch/N_\mathrm{ch}^\mathrm{5\%}$ in Pb+Pb and Xe+Xe collisions. This scaling partially cancels out systematic uncertainties. The error bars represent statistical uncertainties of the measurement, whereas the shaded boxes represent the systematic uncertainties of the data along the $x$- and $y$-axes. The data are compared to predictions from Ref.~\cite{Samanta:2023kfk}, where the estimated geometrical component is also shown.}
\label{fig:fig3}
\end{figure}

To better illustrate these non-monotonic behaviors in UCC, Figure~\ref{fig:fig3} presents $\lr{[\pT]}$ and the normalized observables from Eq.~(\ref{eq:ck2}), each scaled by their respective values at 5\% centrality. The observables are then plotted as a function of $\Nch/N_\mathrm{ch}^\mathrm{5\%}$, $\Nch$ normalized by its value at 5\% centrality, allowing for comparing the two systems on a similar scale along the $x$-axis.

Qualitatively similar behaviors are observed across all observables in both systems, though the variations are slightly weaker in Xe+Xe collisions. This outcome is expected: the smaller mass number of Xe compared to Pb, yields a broader distribution of $\Nch/N_\mathrm{ch}^\mathrm{5\%}$ in Xe+Xe collisions. This broader distribution leads to a weaker suppression of the geometrical component, demonstrating the value of comparing data from collisions involving nuclei of different sizes. It is worth noting that the 8\% higher $\sqn$ in Xe+Xe compared to Pb+Pb collisions is estimated to increase $\Nch$ only by 2.5\%~\cite{ALICE:2018cpu}, which should have negligible effects on the scaled quantities in Figure~\ref{fig:fig3}.

Recent studies~\cite{Samanta:2023amp,Samanta:2023kfk} have modeled $P([\pT])$ as a 2D Gaussian function of $\Nch$ and impact parameter, where the fluctuations of $[\pT]$ at a given $\Nch$ are driven solely by the variations in the impact parameter and $R$. In this model, the increase in $\lr{[\pT]}$ arises from enhanced intrinsic fluctuations at fixed $R$, whereas $k_2$ arises from both geometrical and intrinsic contributions. The $k_3$ originates from a geometrical contribution and a cross-term between geometrical and intrinsic components but has no contribution from pure intrinsic component~\cite{Samanta:2023kfk}. Figure~\ref{fig:fig3} compares predictions from this model to the Pb+Pb data. The model reasonably captures the increase in $\lr{[\pT]}$ and the decrease in $k_2$. However, the predicted $k_3$ values decrease more steeply with $\Nch/N_\mathrm{ch}^\mathrm{5\%}$, a trend also observed in $\Gamma$. The larger $k_3$ and $\Gamma$ values in the data suggest the need for additional sources of skewness in $P([\pT])$~\cite{Giacalone:2020lbm}. Most variations in $k_2$, $k_3$, and $\Gamma$ can be largely attributed to the geometrical component, as indicated by the dashed lines. Although the separation into the two components is clearly feasible within a given model framework, a more detailed investigation of the sensitivities to model parameters is still required.

\begin{figure}[h!]
\centering
\includegraphics[width=0.7\linewidth]{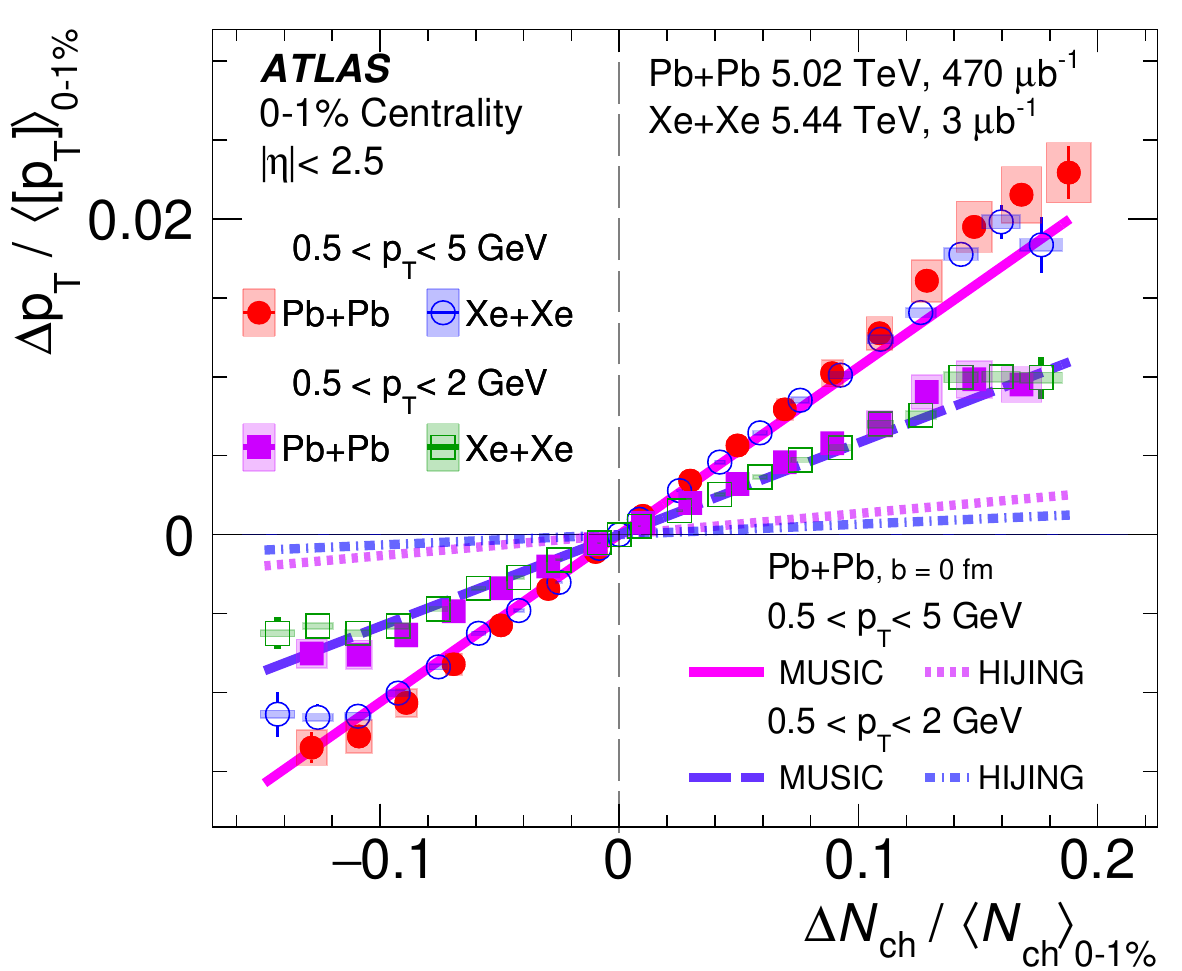}
\caption{ Correlation between $\Delta\pT/\lr{[\pT]}_{0-1\%}$ and $\Delta\Nch/\lr{\Nch}_{0-1\%}$ in the 0--1\% most central Pb+Pb and Xe+Xe collisions for two $\pT$ ranges. The error bars represent the statistical uncertainties of the measurement, whereas the shaded boxes represent the systematic uncertainties of the data along the $x$- and $y$-axes. The data are compared to the MUSIC hydrodynamic model with $c_s^2\approx0.23$ at $T_{\mathrm{eff}}\approx 222$ MeV and the $\HIJING$ model, both at zero impact parameter ($b=0$~fm)~\cite{Samanta:2023amp}.}
\label{fig:fig4}
\end{figure}

For a more direct study of the correlation between $[\pT]$ and $\Nch$ in UCC, a detailed analysis of the 0--1\% most central events is performed. The average values of $[\pT]$ and $\Nch$ for these events are denoted as $\lr{[\pT]}_{0-1\%}$ and $\lr{\Nch}_{0-1\%}$, respectively. Next, $\lr{[\pT]}$ is calculated by averaging $[\pT]$ over events within narrow $\Nch$ slices to obtain $\Delta \pT/\lr{[\pT]}_{0-1\%}$ as a function of $\Delta\Nch/\lr{\Nch}_{0-1\%}$. Here, $\Delta \pT = \lr{[\pT]} - \lr{[\pT]}_{0-1\%}$ and $\Delta\Nch = \Nch - \lr{\Nch}_{0-1\%}$. Figure~\ref{fig:fig4} shows $\Delta \pT/\lr{[\pT]}_{0-1\%}$ plotted against $\Delta\Nch/\lr{\Nch}_{0-1\%}$ for two $\pT$ ranges. A nearly linear relationship is observed, with similar slopes in both systems. However, the slope varies depending on the selected $\pT$ range, indicating a kinematic sensitivity to radial flow effects. Compared to Figure~\ref{fig:fig3}, the better consistency between the two systems can be attributed to the more restrictive centrality range used for normalization in both $x$- and $y$-axes.

The Pb+Pb data are compared to two models: the $\HIJING$ model~\cite{Gyulassy:1994ew}, which lacks final-state interactions, and the state-of-the-art MUSIC model~\cite{Schenke:2010nt}, which incorporates the full hydrodynamic response of the QGP to its initial-state geometry. The $\HIJING$ model grossly underpredicts the slope observed in the data, whereas the MUSIC model reproduces the slopes across both $\pT$ ranges. This finding indicates that the slopes reflect the hydrodynamic response of the QGP in UCC, where the initial transverse size is fixed, but the energy density varies strongly.

A recent study~\cite{Gardim:2019brr} connected the increase of $[\pT]$ in UCC to the speed of sound of the QGP~\cite{Gardim:2019xjs}, calculated as $c_s^2 (T) = d\ln T/d\ln s$, where $T$ and $s$ are the medium's temperature and entropy density, respectively. Since $T$ evolves throughout the QGP's lifetime, $c_s^2$ was estimated at an effective temperature $T_{\mathrm{eff}}$,  approximately 1/3 of the average $\pT$ calculated for all particles~\cite{Gardim:2019xjs, Gardim:2019brr}:
\begin{align}\nonumber
& c_s^2 (T_{\mathrm{eff}}) \propto \frac{d\ln(\lr{[\pT]})}{d\ln(\Nch)} \approx \frac{\Delta\pT/\lr{[\pT]}}{\Delta\Nch/\lr{\Nch}}\;.
\end{align}
According to this model, the measured slope in Figure~\ref{fig:fig4} can be used to estimate $c_s^2(T_{\mathrm{eff}})$. The MUSIC model achieves reasonable agreement with the Pb+Pb data, including its $\pT$ dependence, by using $c_s^2\approx0.23$ with $T_{\mathrm{eff}}\approx 222$ MeV~\cite{Gardim:2019xjs}, values consistent with those reported by the CMS Collaboration~\cite{CMS:2024sgx}. However, the extraction of the $c_s^2$ was shown to be sensitive to several factors, including the kinematic selection of the particles used to define the centrality and $\lr{[\pT]}$~\cite{Nijs:2023bzv,SoaresRocha:2024drz}. Additionally, because the slopes in Figure~\ref{fig:fig4} are driven by the intrinsic component of $P([\pT])$, which can be independently constrained using the higher-order moments in Figure~\ref{fig:fig3}, it is crucial that any model aiming to extract $c_s^2$ can describe these observables simultaneously.

Understanding the initial-state geometry of the QGP and how it drives the hydrodynamic response is a key objective in high-energy nuclear physics. This can be pursued by analyzing the moments of event-by-event transverse momentum distribution $P([\pT])$. This Letter presents the first attempt to disentangle geometrical and intrinsic fluctuations by examining the mean, variance, and skewness of $P([\pT])$ in $^{208}$Pb+$^{208}$Pb and $^{129}$Xe+$^{129}$Xe collisions at $\sqn=5.02$ TeV and 5.44 TeV, respectively. Across a wide $\Nch$ range, the variance and skewness exhibit an approximate power-law scaling consistent with expectations from an independent superposition scenario. However, in ultra-central collisions, all observables deviate from this scaling: the mean shows a distinctive rise, the variance sharply decreases, and the skewness exhibits an increase followed by a sharp decrease. Moreover, the linear rise in $\lr{[\pT]}$ with increasing $\Nch$ is reproduced by a hydrodynamic model using $c_s^2\approx0.23$ at an effective temperature $T_{\mathrm{eff}}\approx 222$ MeV. The centrality dependence of these observables in ultra-central collisions is slightly weaker in the smaller Xe+Xe collisions, and studies involving even smaller nuclei could probe the system size dependence further. Investigating $[\pT]$ fluctuations offers a valuable tool for constraining the initial-state fluctuations and the hydrodynamic response in heavy-ion collisions.
\section*{Acknowledgements}
%

%
%

%
%

We thank CERN for the very successful operation of the LHC and its injectors, as well as the support staff at
CERN and at our institutions worldwide without whom ATLAS could not be operated efficiently.

The crucial computing support from all WLCG partners is acknowledged gratefully, in particular from CERN, the ATLAS Tier-1 facilities at TRIUMF/SFU (Canada), NDGF (Denmark, Norway, Sweden), CC-IN2P3 (France), KIT/GridKA (Germany), INFN-CNAF (Italy), NL-T1 (Netherlands), PIC (Spain), RAL (UK) and BNL (USA), the Tier-2 facilities worldwide and large non-WLCG resource providers. Major contributors of computing resources are listed in Ref.~\cite{ATL-SOFT-PUB-2023-001}.

We gratefully acknowledge the support of ANPCyT, Argentina; YerPhI, Armenia; ARC, Australia; BMWFW and FWF, Austria; ANAS, Azerbaijan; CNPq and FAPESP, Brazil; NSERC, NRC and CFI, Canada; CERN; ANID, Chile; CAS, MOST and NSFC, China; Minciencias, Colombia; MEYS CR, Czech Republic; DNRF and DNSRC, Denmark; IN2P3-CNRS and CEA-DRF/IRFU, France; SRNSFG, Georgia; BMBF, HGF and MPG, Germany; GSRI, Greece; RGC and Hong Kong SAR, China; ISF and Benoziyo Center, Israel; INFN, Italy; MEXT and JSPS, Japan; CNRST, Morocco; NWO, Netherlands; RCN, Norway; MNiSW, Poland; FCT, Portugal; MNE/IFA, Romania; MESTD, Serbia; MSSR, Slovakia; ARRS and MIZ\v{S}, Slovenia; DSI/NRF, South Africa; MICINN, Spain; SRC and Wallenberg Foundation, Sweden; SERI, SNSF and Cantons of Bern and Geneva, Switzerland; MOST, Taipei; TENMAK, T\"urkiye; STFC, United Kingdom; DOE and NSF, United States of America.

Individual groups and members have received support from BCKDF, CANARIE, CRC and DRAC, Canada; CERN-CZ, FORTE and PRIMUS, Czech Republic; COST, ERC, ERDF, Horizon 2020, ICSC-NextGenerationEU and Marie Sk{\l}odowska-Curie Actions, European Union; Investissements d'Avenir Labex, Investissements d'Avenir Idex and ANR, France; DFG and AvH Foundation, Germany; Herakleitos, Thales and Aristeia programmes co-financed by EU-ESF and the Greek NSRF, Greece; BSF-NSF and MINERVA, Israel; NCN and NAWA, Poland; La Caixa Banking Foundation, CERCA Programme Generalitat de Catalunya and PROMETEO and GenT Programmes Generalitat Valenciana, Spain; G\"{o}ran Gustafssons Stiftelse, Sweden; The Royal Society and Leverhulme Trust, United Kingdom.

In addition, individual members wish to acknowledge support from Armenia: Yerevan Physics Institute (FAPERJ); CERN: European Organization for Nuclear Research (CERN PJAS); Chile: Agencia Nacional de Investigaci\'on y Desarrollo (FONDECYT 1230812, FONDECYT 1230987, FONDECYT 1240864); China: Chinese Ministry of Science and Technology (MOST-2023YFA1605700), National Natural Science Foundation of China (NSFC - 12175119, NSFC 12275265, NSFC-12075060); Czech Republic: Czech Science Foundation (GACR - 24-11373S), Ministry of Education Youth and Sports (FORTE CZ.02.01.01/00/22\_008/0004632), PRIMUS Research Programme (PRIMUS/21/SCI/017); EU: H2020 European Research Council (ERC - 101002463); European Union: European Research Council (ERC - 948254, ERC 101089007), Horizon 2020 Framework Programme (MUCCA - CHIST-ERA-19-XAI-00), European Union, Future Artificial Intelligence Research (FAIR-NextGenerationEU PE00000013), Italian Center for High Performance Computing, Big Data and Quantum Computing (ICSC, NextGenerationEU); France: Agence Nationale de la Recherche (ANR-20-CE31-0013, ANR-21-CE31-0013, ANR-21-CE31-0022), Investissements d'Avenir Labex (ANR-11-LABX-0012); Germany: Baden-Württemberg Stiftung (BW Stiftung-Postdoc Eliteprogramme), Deutsche Forschungsgemeinschaft (DFG - 469666862, DFG - CR 312/5-2); Italy: Istituto Nazionale di Fisica Nucleare (ICSC, NextGenerationEU); Japan: Japan Society for the Promotion of Science (JSPS KAKENHI JP22H01227, JSPS KAKENHI JP22H04944, JSPS KAKENHI JP22KK0227, JSPS KAKENHI JP23KK0245); Netherlands: Netherlands Organisation for Scientific Research (NWO Veni 2020 - VI.Veni.202.179); Norway: Research Council of Norway (RCN-314472); Poland: Polish National Agency for Academic Exchange (PPN/PPO/2020/1/00002/U/00001), Polish National Science Centre (NCN 2021/42/E/ST2/00350, NCN OPUS nr 2022/47/B/ST2/03059, NCN UMO-2019/34/E/ST2/00393, UMO-2020/37/B/ST2/01043, UMO-2021/40/C/ST2/00187, UMO-2022/47/O/ST2/00148, UMO-2023/49/B/ST2/04085); Slovenia: Slovenian Research Agency (ARIS grant J1-3010); Spain: Generalitat Valenciana (Artemisa, FEDER, IDIFEDER/2018/048), Ministry of Science and Innovation (MCIN \& NextGenEU PCI2022-135018-2, MICIN \& FEDER PID2021-125273NB, RYC2019-028510-I, RYC2020-030254-I, RYC2021-031273-I, RYC2022-038164-I), PROMETEO and GenT Programmes Generalitat Valenciana (CIDEGENT/2019/027); Sweden: Swedish Research Council (Swedish Research Council 2023-04654, VR 2018-00482, VR 2022-03845, VR 2022-04683, VR 2023-03403, VR grant 2021-03651), Knut and Alice Wallenberg Foundation (KAW 2018.0157, KAW 2018.0458, KAW 2019.0447, KAW 2022.0358); Switzerland: Swiss National Science Foundation (SNSF - PCEFP2\_194658); United Kingdom: Leverhulme Trust (Leverhulme Trust RPG-2020-004), Royal Society (NIF-R1-231091); United States of America: U.S. Department of Energy (ECA DE-AC02-76SF00515), Neubauer Family Foundation.

%
%


%
%
%
%
%
%
%

%
%
%
%
%
%
%
%
%
%
%

\section{Appendix}\label{sec:supp}
\begin{figure}[h!]
\centering
\includegraphics[width=0.7\linewidth]{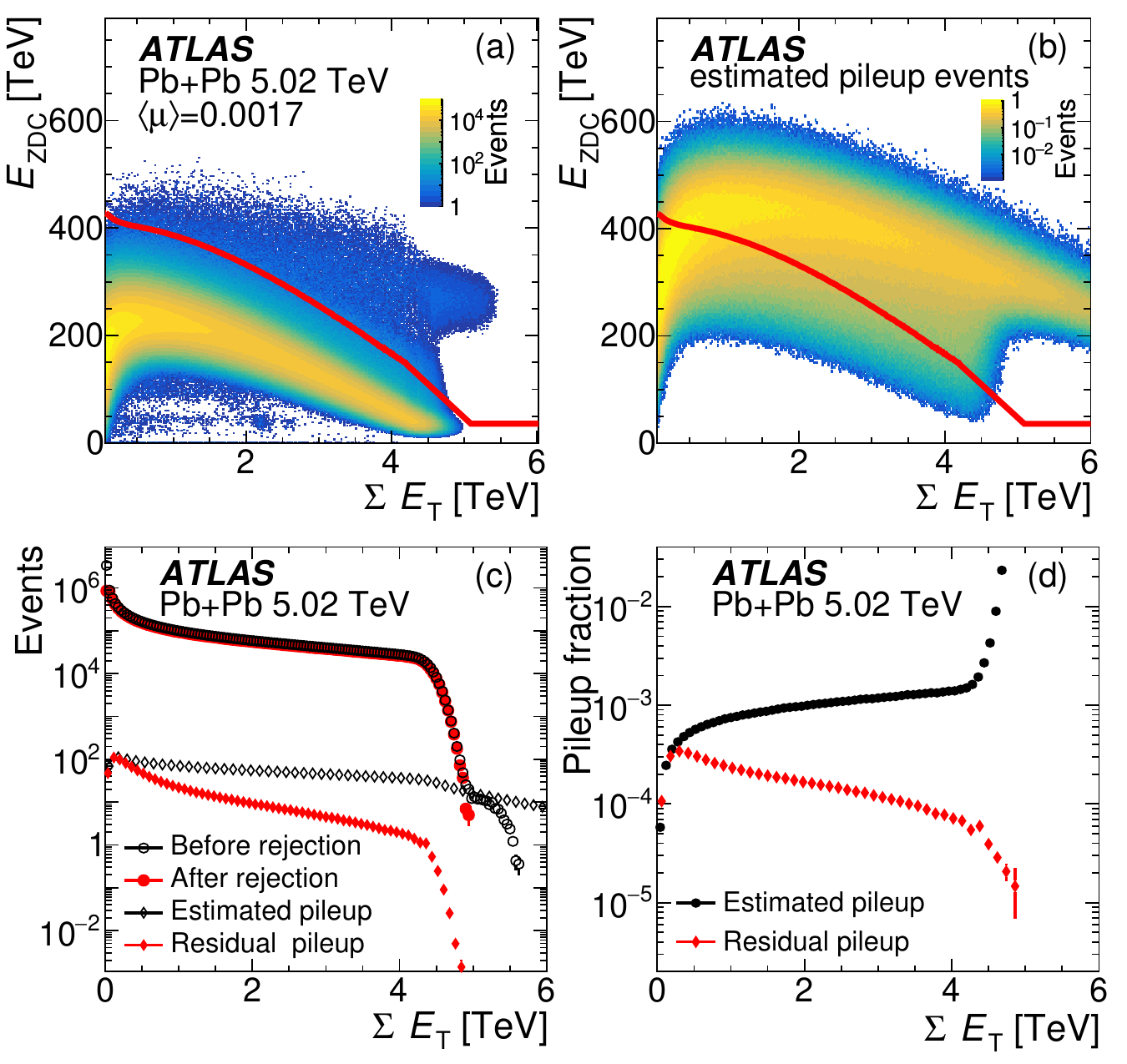}
\caption{ Top: The distributions of energy deposited in the ZDCs vs. FCal for (a) all events, and (b) estimated pileup events, the red line represents the line used for selection of good events. Bottom: As a function $\SumET$, (c) the distributions of all events, pileup events, good events, and residual pileup events, and (d) the fraction of pileup events before and after pileup rejection. The results are obtained for 5.02 TeV Pb+Pb collisions.}
\label{app:1}
\end{figure}

The pileup probability is $\lr{\mu}=0.0017$ in Pb+Pb collisions and 0.00019 in Xe+Xe collisions. Pileup events are not uniformly distributed across $\Nch$ and tend to contribute more significantly in UCC. The impact of pileup events can be estimated and rejected using the anti-correlation between ZDC energy ($E_{\mathrm{ZDC}}$) and FCal $\SumET$ as shown in Figure~\ref{app:1}(a). A typical pileup event in the UCC region consists of a genuine central event with small $E_{\mathrm{ZDC}}$ and large $\SumET$ and a peripheral or mid-central event with large $E_{\mathrm{ZDC}}$ and small $\SumET$, contributing to the satellite band. Good events are selected within a six standard deviation window from the peak value of $E_{\mathrm{ZDC}}$ at a given $\SumET$, as indicated by the red line. Convolving the distribution of good events according to the pileup probability yields an estimate of the pileup event distribution shown in Figure~\ref{app:1}(b).

Figure~\ref{app:1}(c) shows the distributions of all events and estimated pileup events as a function of $\SumET$, along with the distributions of good events and residual pileup events after applying the selection criteria. These criteria significantly reduce the pileup contribution, bringing its fraction down to $\lesssim 0.01\%$ in UCC, as shown in Figure~\ref{app:1}(d). The robustness of the pileup rejection is tested by relaxing the selection criteria, and the results remain largely insensitive to variations in the residual pileup fraction.

\begin{figure}[h!]
\centering
\includegraphics[width=0.8\linewidth]{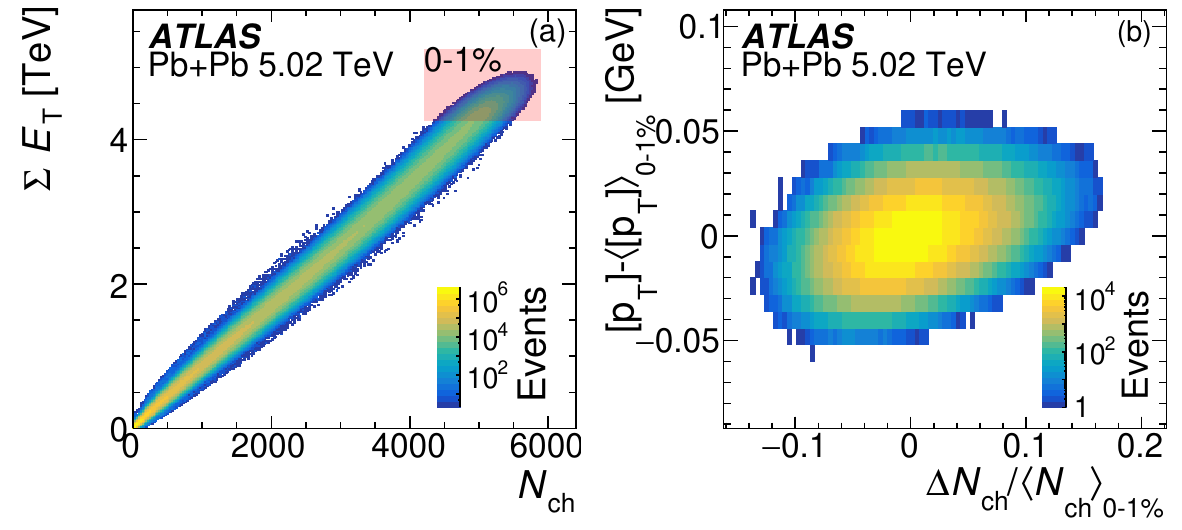}
\caption{(a) Correlation between FCal $\SumET$ and $\Nch$ in Pb+Pb collisions, where the shaded region covers 0--1\% most central events, and (b) correlation between $[\pT]$ and $\Delta\Nch/\lr{\Nch}_{0-1\%}$.}
\label{app:2}
\end{figure}

Figure~\ref{app:2}(a) shows the correlation between $\SumET$ and $\Nch$. The correlation is smeared, implying that the 0--1\% most central events selected based on $\SumET$ span a large range of $\Nch$, as indicated by the shaded box. Consequently, the $[\pT]$ values for these events also span a large range $\Delta\Nch/\lr{\Nch}_{0-1\%}$, as shown by the $x$-axis in Figure~\ref{fig:fig4}.

\begin{figure}[h!]
\centering
\includegraphics[width=0.5\linewidth]{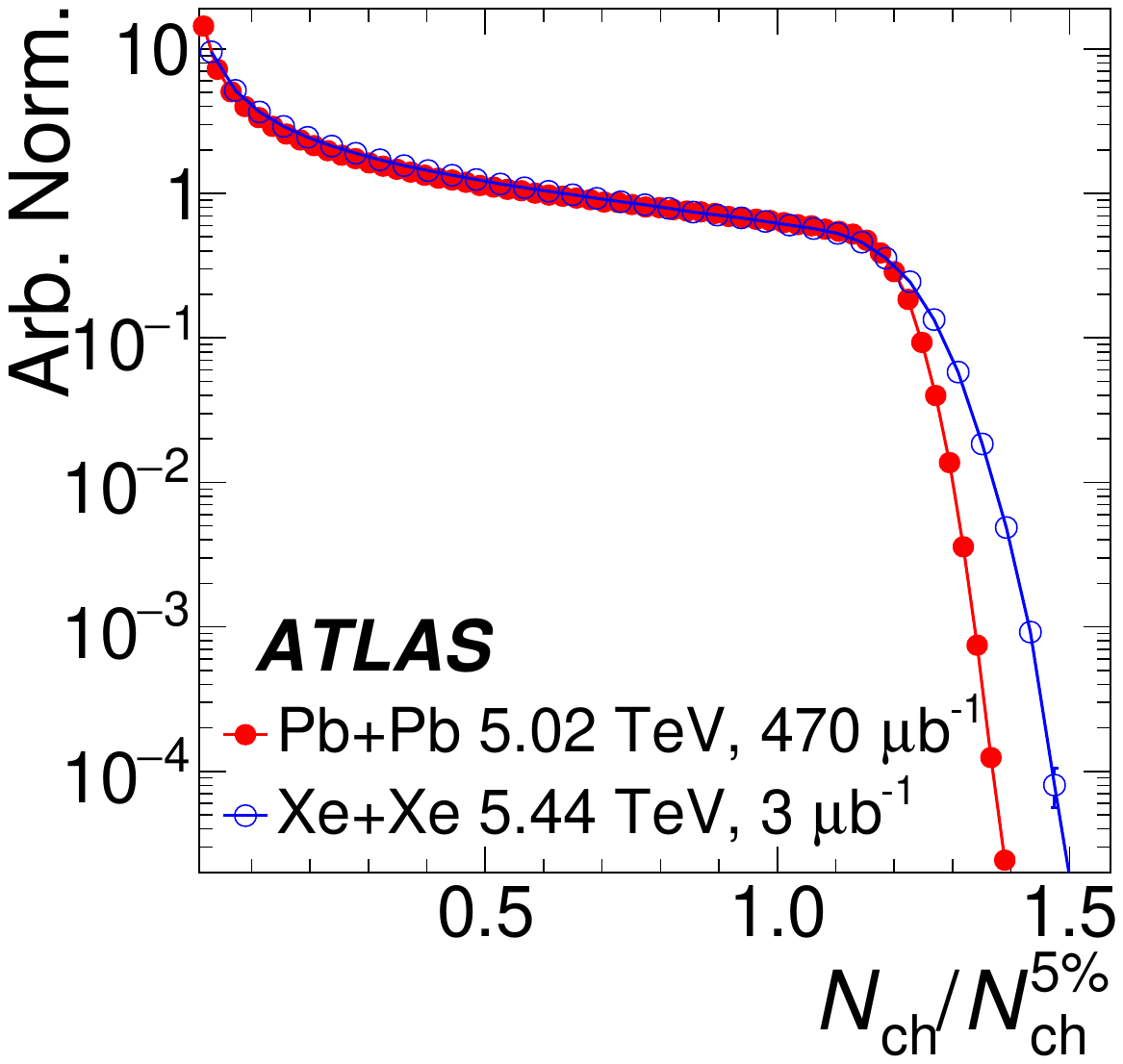}
\caption{The distributions of $\Nch/N^{5\%}_{\mathrm{ch}}$ in Pb+Pb and Xe+Xe collisions.}
\label{fig:app3}
\end{figure}

Figure~\ref{fig:app3} displays the distribution of $\Nch/N^{5\%}_{\mathrm{ch}}$ in Pb+Pb and Xe+Xe collisions, showing the distribution is broader in the smaller Xe+Xe system.

\begin{figure}[h!]
\centering
\includegraphics[width=0.8\linewidth]{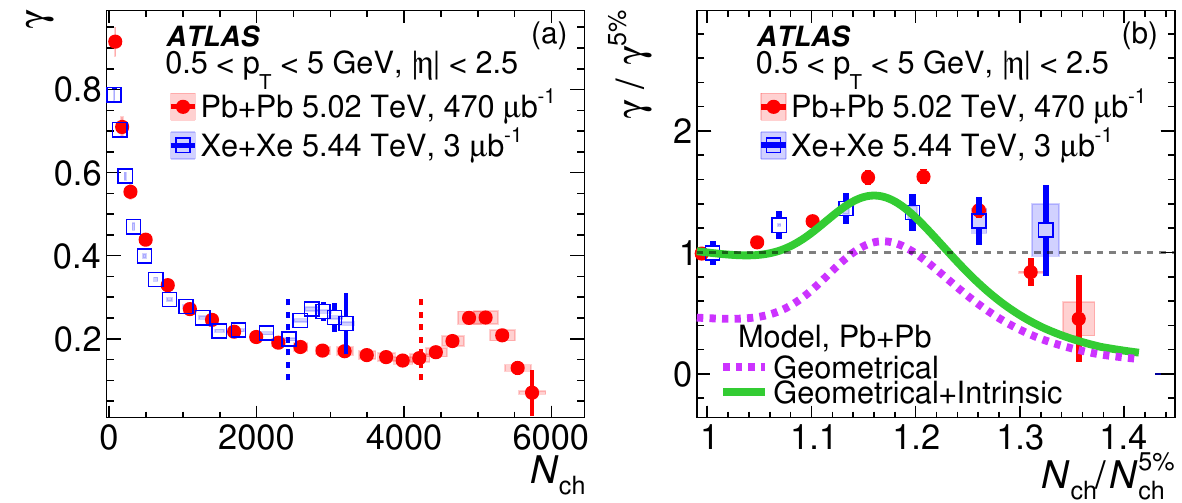}
\caption{(a) $\gamma$ as a function of $\Nch$ and (b) scaled-$\gamma$ as a function of $\Nch / N_\mathrm{ch}^\mathrm{5\%}$ in Pb+Pb and Xe+Xe collisions compared to predictions from Ref.~\cite{Samanta:2023kfk}. The error bars and shaded area represent statistical and systematic uncertainties, respectively. The vertical dashed lines in (a) mark the $\Nch$ values corresponding to 5\% centrality in Pb+Pb and Xe+Xe collisions.}\label{fig:app4}
\end{figure}

Figure~\ref{fig:app4} displays the multiplicity dependence of $\gamma$ and scaled-$\gamma$ in the two systems. They can be derived from Figures~\ref{fig:fig1} and ~\ref{fig:fig3}, but are shown here for completeness.

\clearpage
\printbibliography
\clearpage

\clearpage
 
\begin{flushleft}
\hypersetup{urlcolor=black}
{\Large The ATLAS Collaboration}

\bigskip

\AtlasOrcid[0000-0002-6665-4934]{G.~Aad}$^\textrm{\scriptsize 104}$,
\AtlasOrcid[0000-0001-7616-1554]{E.~Aakvaag}$^\textrm{\scriptsize 17}$,
\AtlasOrcid[0000-0002-5888-2734]{B.~Abbott}$^\textrm{\scriptsize 123}$,
\AtlasOrcid[0000-0002-0287-5869]{S.~Abdelhameed}$^\textrm{\scriptsize 119a}$,
\AtlasOrcid[0000-0002-1002-1652]{K.~Abeling}$^\textrm{\scriptsize 56}$,
\AtlasOrcid[0000-0001-5763-2760]{N.J.~Abicht}$^\textrm{\scriptsize 50}$,
\AtlasOrcid[0000-0002-8496-9294]{S.H.~Abidi}$^\textrm{\scriptsize 30}$,
\AtlasOrcid[0009-0003-6578-220X]{M.~Aboelela}$^\textrm{\scriptsize 45}$,
\AtlasOrcid[0000-0002-9987-2292]{A.~Aboulhorma}$^\textrm{\scriptsize 36e}$,
\AtlasOrcid[0000-0001-5329-6640]{H.~Abramowicz}$^\textrm{\scriptsize 154}$
% ...
\bigskip
\\

% ...
$^{17}$Department for Physics and Technology, University of Bergen, Bergen; Norway.\\
% ...
$^{30}$Physics Department, Brookhaven National Laboratory, Upton NY; United States of America.\\
% ...
$^{36}$$^{(a)}$Facult\'e des Sciences Ain Chock, R\'eseau Universitaire de Physique des Hautes Energies - Universit\'e Hassan II, Casablanca;$^{(b)}$Facult\'{e} des Sciences, Universit\'{e} Ibn-Tofail, K\'{e}nitra;$^{(c)}$Facult\'e des Sciences Semlalia, Universit\'e Cadi Ayyad, LPHEA-Marrakech;$^{(d)}$LPMR, Facult\'e des Sciences, Universit\'e Mohamed Premier, Oujda;$^{(e)}$Facult\'e des sciences, Universit\'e Mohammed V, Rabat;$^{(f)}$Institute of Applied Physics, Mohammed VI Polytechnic University, Ben Guerir; Morocco.\\
% ...
$^{45}$Physics Department, Southern Methodist University, Dallas TX; United States of America.\\
% ...
$^{50}$Fakult\"{a}t Physik , Technische Universit{\"a}t Dortmund, Dortmund; Germany.\\
% ...
$^{56}$II. Physikalisches Institut, Georg-August-Universit\"{a}t G\"ottingen, G\"ottingen; Germany.\\
% ...
$^{104}$CPPM, Aix-Marseille Universit\'e, CNRS/IN2P3, Marseille; France.\\
% ...
$^{119}$$^{(a)}$New York University Abu Dhabi, Abu Dhabi;$^{(b)}$United Arab Emirates University, Al Ain; United Arab Emirates.\\
% ...
$^{123}$Homer L. Dodge Department of Physics and Astronomy, University of Oklahoma, Norman OK; United States of America.\\
% ...
$^{154}$Raymond and Beverly Sackler School of Physics and Astronomy, Tel Aviv University, Tel Aviv; Israel.\\
% ...
$^{a}$ Also Affiliated with an institute covered by a cooperation agreement with CERN.\\
$^{b}$ Also at An-Najah National University, Nablus; Palestine.\\
$^{c}$ Also at Borough of Manhattan Community College, City University of New York, New York NY; United States of America.\\
$^{d}$ Also at Center for Interdisciplinary Research and Innovation (CIRI-AUTH), Thessaloniki; Greece.\\
$^{e}$ Also at Centro Studi e Ricerche Enrico Fermi; Italy.\\
$^{f}$ Also at CERN, Geneva; Switzerland.\\
% ...

\end{flushleft}


\end{document}